\documentclass[journal]{IEEEtran}
\IEEEoverridecommandlockouts
\usepackage{cite}
\usepackage[colorlinks=true,linkcolor=black,anchorcolor=black,citecolor=black,filecolor=black,menucolor=black,runcolor=black,urlcolor=magenta]{hyperref}
\usepackage{amsmath,amssymb,amsfonts,mathrsfs}
\usepackage{algorithmic}
\usepackage{graphicx}
\usepackage{textcomp}
\usepackage{xcolor}
\usepackage{booktabs}
\usepackage[defaultcolor=red,final]{changes}
\def\BibTeX{{\rm B\kern-.05em{\sc i\kern-.025em b}\kern-.08em
    T\kern-.1667em\lower.7ex\hbox{E}\kern-.125emX}}

\ifCLASSOPTIONcompsoc
\usepackage[caption=false,font=normalsize,labelfont=sf,textfont=sf]{subfig}
\else
\usepackage[caption=false,font=footnotesize]{subfig}
\fi

\newcommand{\RE}[1]{\operatorname{Re}\left({#1}\right)}
\newcommand{\IM}[1]{\operatorname{Im}\left({#1}\right)}

\newcommand{\mat}[1]{\boldsymbol{#1}}

\newcommand{\vc}[1]{\mathrm{vec}\left({#1}\right)}

\begin{document}
\bstctlcite{Settings}

\title{Propagation of Linear Uncertainties through \\Multiline Thru-Reflect-Line Calibration}

\author{%
	\IEEEauthorblockN{%
		Ziad~Hatab,~\IEEEmembership{Student~Member,~IEEE,}
		Michael~Ernst~Gadringer,~\IEEEmembership{Senior~Member,~IEEE,}\\ and~Wolfgang~Bösch,~\IEEEmembership{Fellow,~IEEE}
	}%
\thanks{Software implementation and measurements are available online:\\ \url{https://github.com/ZiadHatab/uncertainty-multiline-trl-calibration}}
}%

\markboth{This work has been accepted for publication in the IEEE Transactions on Instrumentation and Measurement}{}

\maketitle

\begin{abstract}
This study proposes a linear approach for propagating uncertainties in the multiline thru-reflect-line (TRL) calibration method for vector network analyzers. The multiline TRL formulation we are proposing applies the law of uncertainty propagation as outlined in the ISO Guide to the Expression of Uncertainty in Measurement (GUM) to both measurement and model uncertainties. In addition, we conducted a Monte Carlo analysis using a combination of measured and synthetic data to model various uncertainties, such as measurement noise, reflect asymmetry, line mismatch, and line length offset. The results of our linear uncertainty formulation demonstrate agreement with the Monte Carlo analysis and provide a more efficient means of assessing the uncertainty budget of the multiline TRL calibration.
\end{abstract}


\section{Introduction}
\label{sec:1}
\IEEEPARstart{T}{he} calibration of vector network analyzers (VNAs) is crucial to establish a reference measurement plane for the measured device under test (DUT). However, random measurement variations can arise from various sources, such as instrumentation or calibration standards. To account for these uncertainties, it is essential to quantify them at the new measurement plane through uncertainty propagation.

A common method for uncertainty propagation is to consider disturbances as multivariate Gaussian distributions and to use the first-order Taylor series expansion of the underlying mathematical model used in the calibration process. This approach is known as the law of uncertainty propagation, outlined in the ISO Guide to the Expression of Uncertainty in Measurement (GUM) \cite{GuidesinMetrology2011}. However, the law of uncertainty propagation can only be applied if the sources of uncertainties can be explicitly expressed in the mathematical model used for calibration. For example, the short-open-load-thru (SOLT) calibration technique relates directly to all calibration standards, and the law of uncertainty propagation can be applied without issues. However, for VNA self-calibration methods, such as multiline thru-reflect-line (TRL), partially defined calibration standards present a challenge. The lack of a direct relationship between the calibration equations and the standards makes it difficult to assess uncertainties using the law of uncertainty propagation directly.

Uncertainty propagation for classical TRL calibration has been studied in many publications \cite{Hall2018,Zhao2020,Garelli2012,Wollensack2017}. In terms of multiline TRL calibration, uncertainty propagation has been addressed by the National Institute of Standards and Technology (NIST) in their StatistiCAL Plus and Microwave Uncertainty Framework (MUF) software packages, where the multiline TRL algorithm is based on optimization procedures \cite{Williams2003,Williams2003a,Williams2011}. Other approaches to uncertainty evaluation in multiline TRL are via Monte Carlo (MC) analysis \cite{Luan2020}, where the classical multiline TRL algorithm \cite{Marks1991,DeGroot2002} is treated as a black box. Such MC simulations are usually time intensive. The challenges of applying the law of uncertainty propagation to multiline TRL calibration arise from the mathematical limitations of the classical algorithm \cite{Marks1991,DeGroot2002}. The multiline TRL algorithm involves perturbing the solutions of TRL pairs by selecting a common line as the reference for all line pairs. However, this heuristic approach of selecting a common line can result in inconsistencies across the frequency axis.

In a prior publication, we proposed a unique mathematical formulation for multiline TRL calibration that simplifies the process by reducing it to solving a single $4\times4$ eigenvalue problem, regardless of the number of line standards used \cite{Hatab2022}. We have demonstrated that our method is more reliable than previous approaches by considering all line pairs and utilizing optimal weighting to the measurements, rather than weighting the eigenvectors of the TRL pairs. Furthermore, in \cite{Hatab2022b}, we demonstrated the application of linear uncertainty propagation in multiline TRL calibration. The objective of this article is to provide more details on linear uncertainty propagation in multiline TRL calibration, which was briefly discussed in \cite{Hatab2022b}. For this purpose, we have refined our original formulation by relying exclusively on matrix decomposition, resulting in a more efficient and direct application of the law of uncertainty propagation. Additionally, we address the update of the measurement covariance matrix to account for different sources of uncertainty, such as line mismatch, that cannot be explicitly expressed in the calibration.

The remaining of this article is structured as follows. Section \ref{sec:2} presents the mathematical formulation of multiline TRL calibration, incorporating a new weighting matrix technique based on Takagi decomposition. In Section \ref{sec:3}, we provide a comprehensive framework for handling various types of uncertainties in linear uncertainty propagation. Section \ref{sec:4} compares our method with the MC method using on-wafer measurements and synthetic data. Finally, we conclude in Section \ref{sec:5}.

\section{Formulation of Multiline TRL Calibration}
\label{sec:2}

\begin{figure}[!ht]
	\centering
	\includegraphics[width=0.95\linewidth]{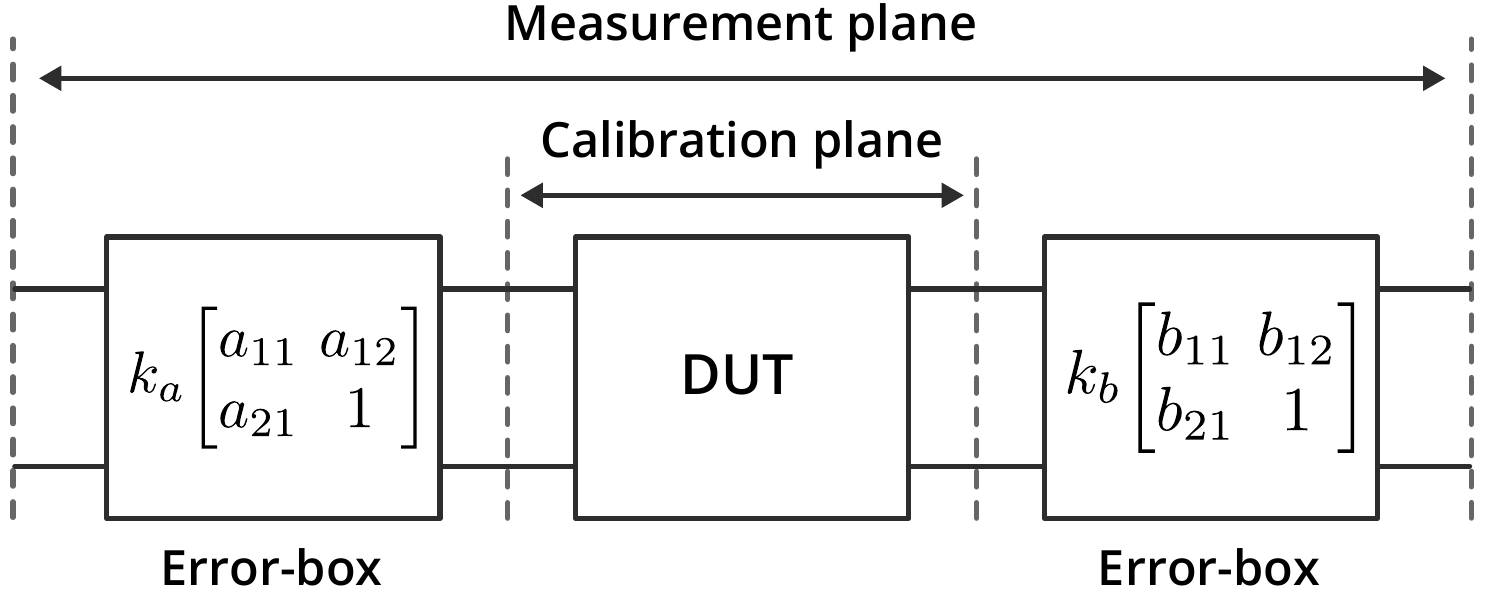}
	\caption{Two-port VNA error-box model.}
	\label{fig:2.1}
\end{figure}

A two-port VNA can be modeled with the error-box model \cite{Marks1997}, as depicted in Fig \ref{fig:2.1}. Mathematically, we describe the error-box model with T-parameters as seven error terms,
\begin{equation}
	\mat{M}_\mathrm{dut} = \underbrace{k_ak_b}_{k}\underbrace{\left[\begin{matrix}a_{11} & a_{12}\\a_{21} & 1\end{matrix}\right]}_{\mat{A}}\mat{T}_\mathrm{dut} \underbrace{\left[\begin{matrix}b_{11} & b_{12}\\b_{21} & 1\end{matrix}\right]}_{\mat{B}},
	\label{eq:2.1}
\end{equation}
where $\mat{M}_\mathrm{dut}$ and $\mat{T}_\mathrm{dut}$ are the T-parameters of the uncalibrated measurement and the actual DUT, respectively. The matrices $\mat{A}$ and $\mat{B}$ are the scaled T-parameters of the error-boxes holding the first six error terms, and $k$ is the 7th error term. 

In contrast to the conventional formulation of the error-box model in (\ref{eq:2.1}), we use a matrix vectorization and the Kronecker product formulation and their relationship \cite{Brewer1978} to redefine the problem as
\begin{equation}
	\vc{\mat{M}_\mathrm{dut}} = k(\mat{B}^T\otimes\mat{A})\vc{\mat{T}_\mathrm{dut}},
	\label{eq:2.2}
\end{equation}
where $\otimes$ is the Kronecker product and $\vc{\cdot}$ is the vectorization operation (more details on these operations, see \cite{Brewer1978}). 

The transmission line standards used in multiline TRL are of different lengths with the same cross-section geometry. Therefore, when measuring a line standard, the T-parameters are given by
\begin{equation}
	\mat{L}_i = \left[\begin{matrix} e^{- \gamma l_{i}} & 0\\0 & e^{\gamma l_{i}} \end{matrix}\right],
	\label{eq:2.3}
\end{equation}
where $\gamma$ is the propagation constant of the line standards (the same for all line standards) and $l_i$ is the length of the $i$th line standard. Assuming that we have $N\geq2$ transmission line standards (including the thru standard), the primary equations from \cite{Hatab2022} we care about are
\begin{subequations}
	\begin{align}
		\mat{M} &= k\mat{X}\mat{L},\label{eq:2.4a}\\[5pt] 
		\mat{D}^{-1}\mat{M}^{T}\mat{P}\mat{Q} &= \frac{1}{k}\mat{L}^{T}\mat{P}\mat{Q}\mat{X}^{-1},\label{eq:2.4b}
	\end{align}
	\label{eq:3.5}
\end{subequations}
where 
\begin{subequations}
	\begin{align}
		\mat{M} &= \begin{bmatrix} \vc{\mat{M}_1} & \vc{\mat{M}_2} & \cdots & \vc{\mat{M}_N} \end{bmatrix},\label{eq:2.5a}\\[5pt]
		\mat{L} &= \begin{bmatrix} \vc{\mat{L}_1} & \vc{\mat{L}_2} & \cdots & \vc{\mat{L}_N} \end{bmatrix},\label{eq:2.5b}\\[5pt]
		\mat{X} &= \mat{B}^T\otimes\mat{A},\label{eq:2.5c}\\[5pt]
		\mat{D} &= \mathrm{diag}\left( \begin{bmatrix} \det(\mat{M}_1) & \cdots & \det(\mat{M}_N) \end{bmatrix}\right),\label{eq:2.5d}\\[5pt]
		\mat{P} &= \begin{bmatrix}
			1 & 0 & 0 & 0\\
			0 & 0 & 1 & 0\\
			0 & 1 & 0 & 0\\
			0 & 0 & 0 & 1
		\end{bmatrix}, \quad \mat{Q} = \begin{bmatrix}
			0 & 0  & 0  & 1\\
			0 & -1 & 0  & 0\\
			0 & 0  & -1 & 0\\
			1 & 0  & 0  & 0
		\end{bmatrix}.\label{eq:2.5e}
	\end{align}
	\label{eq:2.5}
\end{subequations}

\subsection{Construction of the eigenvalue problem}
The eigenvalue problem is formulated by multiplying an $N\times N$ weighting matrix $\mat{W}$ on the right-hand side of (\ref{eq:2.4a}). Then, we multiply the new equation on the left-hand side of (\ref{eq:2.4b}). This finally results in
\begin{equation}
	\underbrace{\mat{M}\mat{W}\mat{D}^{-1}\mat{M}^T\mat{P}\mat{Q}}_{\mat{F}:\ 4\times4} = \mat{X}\underbrace{\mat{L}\mat{W}\mat{L}^T\mat{P}\mat{Q}}_{\mat{H}:\ 4\times4}\mat{X}^{-1}.
	\label{eq:2.6}
\end{equation}

In general, the problem presented in (\ref{eq:2.6}) is a similarity equation between the matrices $\mat{F}$ and $\mat{H}$, where $\mat{X}$ is the transformation matrix. However, choosing an appropriate $\mat{W}$, the matrix $\mat{H}$ can turn into a diagonal matrix. As a result, the similarity problem in (\ref{eq:2.6}) transforms into an eigenvalue problem. From \cite{Hatab2022}, the optimal $\mat{W}$ that maximizes the eigenvalue separation is found to be:
\begin{equation}
	\mat{W}^H = \mat{z}\mat{y}^T - \mat{y}\mat{z}^T = \begin{bmatrix}
		\mat{z} & \mat{y}
	\end{bmatrix}\left[\begin{matrix}
		0 & 1\\ -1 & 0
	\end{matrix}\right]\begin{bmatrix}
		\mat{z}^T \\ \mat{y}^T
	\end{bmatrix},
	\label{eq:2.7}
\end{equation}
where $(\cdot)^H$ is the Hermitian transpose (conjugate transpose). The vectors $\mat{y}$ and $\mat{z}$ are given by
\begin{subequations}
	\begin{align}	
		\mat{y} &= \begin{bmatrix}
			e^{\gamma l_1} & \cdots & e^{\gamma l_N}
		\end{bmatrix}^T,\label{eq:2.8a}\\[5pt]
		\mat{z} &= \left[\begin{matrix} 
			e^{-\gamma l_1} &  \cdots & e^{-\gamma l_N}
		\end{matrix}\right]^T.\label{eq:2.8b}
	\end{align}
	\label{eq:2.8}
\end{subequations}

Therefore, under the condition that $\mat{W}$ was chosen according to (\ref{eq:2.7}), the eigenvalue problem is now given by
\begin{equation}
	\mat{F} = \mat{X}\begin{bmatrix}
		-\lambda & 0 & 0 & 0\\
		0 & 0 & 0 & 0\\
		0 & 0 & 0 & 0\\
		0 & 0 & 0 & \lambda
	\end{bmatrix}\mat{X}^{-1},
	\label{eq:2.9}
\end{equation}
with 
\begin{equation}
	\lambda = \sum\limits_{\begin{gathered}\\[-18pt]
			\scriptstyle i=1\\[-8pt]\scriptstyle i<j\leq N\end{gathered}}^{N-1}\left|e^{\gamma (l_i-l_j)}-e^{-\gamma (l_i-l_j)}\right|^2.
	\label{eq:2.10}
\end{equation}

The calibration coefficients can be solved as the eigenvectors associated with the eigenvalues $\pm\lambda$. However, eigenvector solutions are only known up to a complex scalar factor. Therefore, the calibration coefficients obtained from the eigenvectors must be normalized to obtain a unique solution. Additionally, as $\mat{X}$ is constructed as a Kronecker product, it can be shown that the remaining inner columns of $\mat{X}$ can be obtained directly from the normalized eigenvectors of $\pm\lambda$ \cite{Hatab2022}. The denormalization of the eigenvectors is accomplished with the measurements of the reflect and the thru standards, while simultaneously solving for the 7th error term $k$. The equations for the denormalization can be found in \cite{Hatab2022}. Lastly, the propagation constant can be determined as a final step from the normalized calibration coefficients and the measurements of the line standards. In this step, the exponential terms of the line standards are extracted and combined in the least squares sense. This estimate of the propagation constant can be used to shift the reference plan if desired or used in other measurements such as characteristic impedance estimation \cite{Marks1991a} or permittivity measurement \cite{Hatab2022a}.

\subsection{Computing the weighting matrix}
The weighting matrix $\mat{W}$ is an essential part of achieving a reliable calibration, as it optimally weights the measurement pairs according to their importance. Measurement pairs that are near singular are given less weight. In general, we can determine $\mat{W}$ by knowing the propagation constant and the length of the line standards, as given in (\ref{eq:2.7}). In \cite{Hatab2022}, we originally proposed to compute the propagation constant from the measurements through non-linear optimization. However, numerical optimization procedures can pose issues in linear uncertainty propagation, as such methods are solved iteratively, and a proper convergence is required. 

In this article, we propose a new technique for deriving $\mat{W}$ without having to compute the propagation constant. We first define a new equation by multiplying  (\ref{eq:2.4a}) with the right-hand side of (\ref{eq:2.4b}). In this way, we get the following equation:
\begin{equation}
	\underbrace{\mat{D}^{-1}\mat{M}^{T}\mat{P}\mat{Q}\mat{M}}_{\text{measurements: }N\times N} = \underbrace{\mat{L}^{T}\mat{P}\mat{Q}\mat{L}}_{\text{model: }N\times N},
	\label{eq:2.11}
\end{equation}
where 
\begin{equation}
	\mat{L}^{T}\mat{P}\mat{Q}\mat{L} = \mat{z}\mat{y}^T + \mat{y}\mat{z}^T = \begin{bmatrix}
		\mat{z} & \mat{y}
	\end{bmatrix}\left[\begin{matrix}
		0 & 1\\ 1 & 0
	\end{matrix}\right]\begin{bmatrix}
		\mat{z}^T \\ \mat{y}^T
	\end{bmatrix}.
	\label{eq:2.12}
\end{equation}
with $\mat{y}$ and $\mat{z}$ being the same as defined in (\ref{eq:2.8}). 

From (\ref{eq:2.11}), it is clear that the equation does not depend on error-boxes. Additionally, from (\ref{eq:2.12}), we can see that its structure is very similar to the equation for $\mat{W}^H$ given in (\ref{eq:2.7}). The only difference between (\ref{eq:2.7}) and (\ref{eq:2.12}) is that the subtraction changes to a summation sign. This structural similarity implies that both equations share the same vector basis. To derive this vector basis from (\ref{eq:2.11}) and construct $\mat{W}$ accordingly, we use the Eckart-Young-Mirsky theorem \cite{Eckart1936,Mirsky1960} by performing a singular value decomposition (SVD) and recovering the first two dominant singular values and their associated singular vectors. This process can be written as 
\begin{equation}
	\mat{L}^{T}\mat{P}\mat{Q}\mat{L} = s_1\mat{u}_1\mat{v}_1^H + s_2\mat{u}_2\mat{v}_2^H,
	\label{eq:2.13}
\end{equation}
where $s_i$, $\mat{u}_i$ and $\mat{v}_i$ are the singular values and associated left and right singular vectors of the measurements in (\ref{eq:2.11}). Furthermore, as (\ref{eq:2.11}) has a symmetric structure, we can use the Takagi decomposition to split the matrix into a symmetric product as
\begin{equation}
	s_1\mat{u}_1\mat{v}_1^H + s_2\mat{u}_2\mat{v}_2^H = \mat{U}_{t}\mat{S}_t\mat{U}_t^T = \underbrace{\mat{G}}_{N\times 2}\mat{G}^T,
	\label{eq:2.14}
\end{equation}
where $\mat{U}_t$ is a unitary matrix that holds the Takagi singular vectors and $\mat{S}_t$ is a diagonal matrix with positive real-valued singular values (equal to the singular values in SVD). As a result, $\mat{G} = \mat{U}_t\sqrt{\mat{S}_t}$. Takagi decomposition is analogous to the eigen-decomposition of real-valued symmetric matrices but for complex-valued symmetric matrices (without conjugation). In fact, the Takagi decomposition can be computed through SVD or eigen-decomposition \cite{Chebotarev2014,Che2018}. Therefore, we can simultaneously apply the Eckart-Young-Mirsky theorem and perform Takagi decomposition in a single step, either with SVD or eigen-decomposition. 

Lastly, the matrix $\mat{G}$ is the common vector basis between equations (\ref{eq:2.7}) and (\ref{eq:2.12}). As a result, $\mat{W}$ is determined by
\begin{equation}
	\mat{W}^H = \pm \mat{G}\left[\begin{matrix}
		0 & j\\ -j & 0
	\end{matrix}\right]\mat{G}^T,
	\label{eq:2.15}
\end{equation}
where the sign ambiguity results from not knowing the order of the sum from the matrix decomposition. This sign ambiguity can be resolved by choosing the answer which has the smallest Euclidean distance to a known estimate, for example, an estimate extracted from material properties. The complex number $j$ in (\ref{eq:2.15}) is to compensate for the square-root matrix of the permutation matrix in (\ref{eq:2.12}), which is implicit in the matrix $\mat{G}$.

\section{Linear Uncertainty Propagation}
\label{sec:3}
The law of uncertainty propagation described by the ISO GUM \cite{GuidesinMetrology2011} is defined for real-valued quantities. However, S-parameters are complex-valued quantities. Therefore, to express uncertainties in S-parameters, we need to split the complex-valued expression into their real-valued equivalent. Suppose that we have the complex-valued function $\mat{f}(\mat{z}): \mathbb{C}^m\rightarrow\mathbb{C}^n$, we define a real-valued equivalent of $\mat{f}(\mat{z})$ by the function $\mat{h}(\mat{r}): \mathbb{R}^{2m}\rightarrow\mathbb{R}^{2n}$, where
\begin{subequations}
	\begin{align}
		\mat{r} &= \RE{\mat{z}}\otimes\begin{bmatrix} 1\\0 \end{bmatrix}+ \IM{\mat{z}}\otimes\begin{bmatrix} 0\\1 \end{bmatrix},\label{eq:3.1a}\\[5pt]
		\mat{h} &= \RE{\mat{f}}\otimes\begin{bmatrix} 1\\0 \end{bmatrix}+ \IM{\mat{f}}\otimes\begin{bmatrix} 0\\1 \end{bmatrix}.\label{eq:3.1b}
	\end{align}
	\label{eq:3.1}
\end{subequations}

Then, the multidimensional Taylor expansion \cite{Khang2019} of $\mat{h}(\mat{r})$ is given by
\begin{equation}
	\mat{h}(\mat{r}) = \mat{h}(\mat{\mu_r}) + \mat{J}_{\mat{h}}(\mat{\mu_r})\mat{r} + \mathcal{O}(\mat{r}^2),
	\label{eq:3.2}
\end{equation}
where $\mat{\mu_r}$ is the mean value around which the expansion is evaluated, $\mat{J}_{\mat{h}}(\mat{\mu_r})$ is the Jacobian matrix of $\mat{h}(\mat{r})$ evaluated at $\mat{\mu_r}$, and $\mathcal{O}(\mat{r}^2)$ indicates the remaining higher-order terms.

If we assume that the input parameters follow a multivariate Gaussian distribution $\mat{r}\sim\mathcal{N}(\mat{\mu_r},\mat{\Sigma_r})$, where $\mat{\Sigma_r}$ is the covariance matrix of $\mat{r}$, and consider only the first-order Taylor expansion of $\mat{h}(\mat{r})$, then the mean and covariance of the output are also distributed after multivariate Gaussian $\mat{h}\sim\mathcal{N}(\mat{\mu_h},\mat{\Sigma_h})$, where
\begin{subequations}
	\begin{align}
		\mat{\mu_h} &= \mat{h}(\mat{\mu_r}),\label{eq:3.3a}\\[5pt]
		\mat{\Sigma_h} &= \mat{J}_{\mat{h}}(\mat{\mu_r})\mat{\Sigma_r}\mat{J}_{\mat{h}}^T(\mat{\mu_r}).\label{eq:3.3b}
	\end{align}
	\label{eq:3.3}
\end{subequations}

To perform linear uncertainty propagation, it is necessary to evaluate the function and its Jacobian at the mean value of the input parameter. The analytical calculation of the Jacobian can be challenging, particularly for complex operations such as eigenvalue decomposition. However, there are various ways to estimate the Jacobian, such as using software packages that numerically estimate the Jacobian using finite-difference method, or analytically using automatic differentiation (AD) techniques. Examples of software packages that perform AD calculations are \cite{Zeier_2012} and \cite{Hall2022}.

In the following subsections, we discuss the various uncertainty contributions in multiline TRL calibration and how to propagate them linearly through the calibration.

\subsection{Measurement and forward model uncertainty}
\label{sec:3a}
Measurement uncertainties originate from the instrument itself. Such uncertainties could come in the form of additive or multiplicative noise. The general approach to model measurement uncertainties of S-parameters is through sample statistics. The covariance matrix of measured S-parameters can be determined from a sample covariance. If we assume a multivariate Gaussian distribution, then the unbiased sample covariance is given by
\begin{equation}
	\mat{\Sigma_S} \approx \frac{1}{n-1}\sum_{i=1}^{n} (\mat{h}_i-\mat{\mu_h})(\mat{h}_i-\mat{\mu_h})^T,
	\label{eq:3.4}
\end{equation} 
where $n$ is the total number of samples, $\mat{h}_i$ is the vectorized real-valued representation of the measurements of the $i$th S-parameters, and $\mat{\mu_h}$ is the corresponding mean value. $\mat{h}_i$ is defined similarly to (\ref{eq:3.1}),
\begin{equation}
	\mat{h}_i = \RE{\vc{\mat{S}_i}}\otimes\begin{bmatrix} 1\\0 \end{bmatrix} + \IM{\vc{\mat{S}_i}}\otimes\begin{bmatrix} 0\\1 \end{bmatrix}.
	\label{eq:3.5}
\end{equation}

In many modern VNAs, the wave parameters of a two-port device can be directly measured for a given port source. Using the wave parameters, we can compute the S-parameters by combining the measurements of both source directions \cite{Rumiantsev2008}, as stated below, 
\begin{equation}
	\mat{S} = \begin{bmatrix}
		b_{11}^\prime & b_{12}^\prime\\[5pt]
		b_{21}^\prime & b_{22}^\prime
	\end{bmatrix}\begin{bmatrix}
		a_{11}^\prime & a_{12}^\prime\\[5pt]
		a_{21}^\prime & a_{22}^\prime
	\end{bmatrix}^{-1},
	\label{eq:3.6}
\end{equation}
where $a_{ij}^\prime$ and $b_{ij}^\prime$ represent the measured incident and reflected waves at the receiver port $i$ when sourced by port $j$, respectively. The formulation in equation (\ref{eq:3.6}) accounts for switch terms \cite{Marks1997}, eliminating the possibility of correlation between measurements through switch term correction equations.

In the multiline TRL calibration, as we deal with T-parameters, the sample covariance can be propagated after converting from S- to T-parameters. Knowing the covariance of the measurements allows for the direct propagation of uncertainties at every step of the calibration through the chain rule of Jacobian matrices, ensuring accurate and consistent uncertainty calculations throughout the calibration process.

Forward model uncertainties are parameters that are considered during the calculation of calibration coefficients. An example of such a parameter is the length of line standards used to determine the propagation constant. Even though the propagation constant itself is not directly used in the calibration process, it can affect the reference plane by specifying its location. Additionally, when the calibration coefficients are denormalized using reflection measurements, it is possible to include the uncertainty of the unknown symmetric reflect standard by treating it as an independent variable in the denormalization equation. This can be done by substituting its estimated value into the corresponding Jacobian matrix.

\begin{figure}[th!]
	\centering
	\includegraphics[width=.95\linewidth]{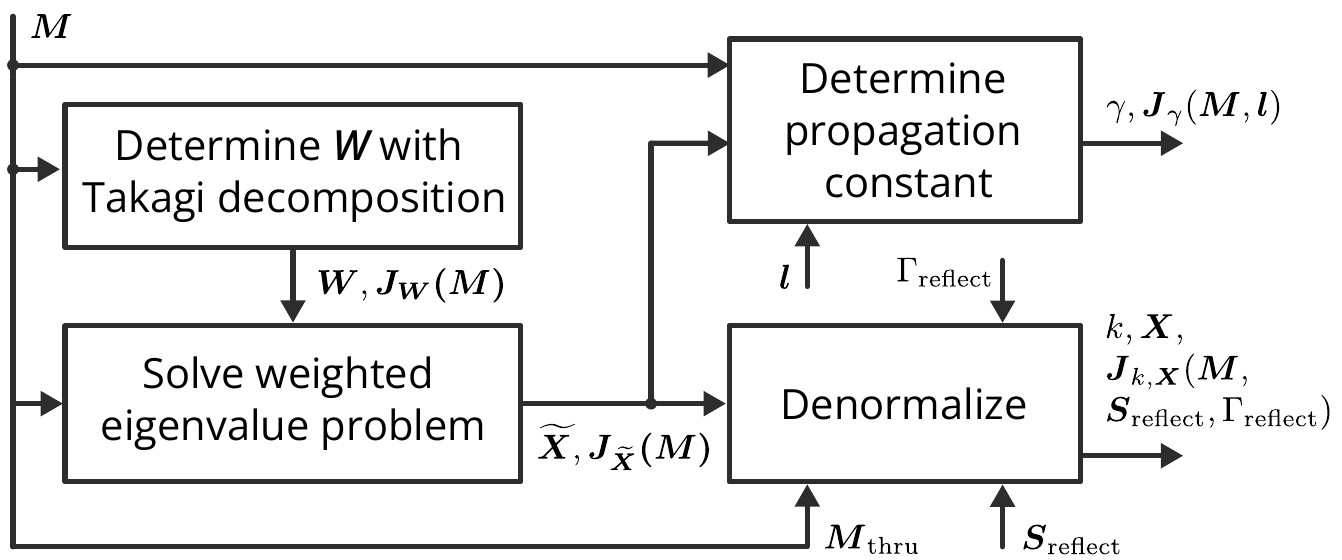}
	\caption{A flow diagram of multiline TRL calibration with forward uncertainty propagation showing the steps involved in the process. It includes the use of T-parameter measurements of all line standards (including thru) in matrix $\mat{M}$, weighting matrix $\mat{W}$, normalized calibration coefficients $\widetilde{\mat{X}}$, a vector of length of all line standards (in reference to the thru standard) $\mat{l}$, S-parameter measurements of the reflect standard $\mat{S}_\mathrm{reflect}$ and its estimated reflection coefficient $\Gamma_\mathrm{reflect}$.}
	\label{fig:3.1}
\end{figure}

The flow diagram of the forward uncertainty propagation in multiline TRL calibration is shown in Fig. \ref{fig:3.1}. The Jacobian matrix is updated at each step, and the covariance of the calibration coefficients is determined using (\ref{eq:3.3b}). If the reference plane is shifted, the Jacobian matrix of the calibration coefficients shall be updated to account for the dependency on the shifted reference plane.

\subsection{Inverse model uncertainty}
\label{sec:3b}

Inverse model uncertainty quantification is the process of identifying the uncertainty contributions arising from the model's limitations used to estimate calibration coefficients. Various sources of uncertainty cannot be forward propagated in the multiline TRL algorithm, such as line mismatch, connecting/probing repeatability, and cable movement. For instance, the eigenvalue problem in multiline TRL assumes that all line standards have the same cross-section (i.e., the same characteristic impedance and propagation constant) and that error-boxes remain unchanged for each measurement. In reality, these assumptions are not always accurate due to randomness. However, on average, they are considered valid if the underlying phenomena are unbiased (i.e., their effects average to zero).

Various techniques can be used to address inverse model uncertainty quantification, such as numerical methods based on Bayesian modeling \cite{Wu2018, Wu2019}. In this work, a more straightforward approach is taken by reformulating the inverse problem as a forward uncertainty propagation by updating the covariance matrix of the measurements. It is important to note that the measurement noise and the model uncertainties are two independent statistical processes. Measurement noise originates from the VNA, while model uncertainties originate from the calibration standards. By assuming that each process follows a zero-mean multivariate Gaussian distribution, the overall distribution of the measurement remains multivariate Gaussian with the same mean value but with a total covariance comprising the sum of the measurement noise and model uncertainties as represented in the equation:
\begin{equation}
	\mat{\Sigma}_\mathrm{M} = \mat{\Sigma}_\mathrm{N} + \mat{\Sigma}_\mathrm{F} + \mat{\Sigma}_\mathrm{I},
	\label{eq:3.7}
\end{equation}
where $\mat{\Sigma}_\mathrm{N}$ is the covariance due to noise, $\mat{\Sigma}_\mathrm{F}$ represents the model uncertainties that can be directly accounted for in the forward uncertainty propagation, and $\mat{\Sigma}_\mathrm{I}$ represents every source of uncertainty that is addressed with inverse uncertainty propagation. In this work, we focus on the line mismatch, as it can be easily modeled in relation to the measurements. Still, the presented technique can also be applied to other uncertainty types, such as cable movement, connection repeatability, or probing uncertainties, if they can be adequately modeled in relation to the measurements.

To quantify the line mismatch uncertainty, we can use the generalized mismatched line model in the error-box model given by
\begin{equation}
	\vc{\mat{M}_i^\prime} = k\mat{X}\vc{L_i^\prime},
	\label{eq:3.8}
\end{equation}
where $\mat{L}_i^\prime$ is the $i$th mismatched line model given by \cite{Marks1992},
\begin{equation}
	\mat{L}_i^\prime = \frac{1}{1-\Gamma_i^2}\begin{bmatrix}
		1 & \Gamma_i\\
		\Gamma_i & 1
	\end{bmatrix}\begin{bmatrix}
		e^{-\gamma_i l_i} & 0\\
		0 & e^{\gamma_i l_i}
	\end{bmatrix}\begin{bmatrix}
		1 & -\Gamma_i\\
		-\Gamma_i & 1
	\end{bmatrix},
	\label{eq:3.9}
\end{equation}
with $\Gamma_i = \delta_{Z_i}/(2\mu_Z + \delta_{Z_i})$ and $\gamma_i = \mu_{\gamma} + \delta_{\gamma_i}$, where $\delta_{Z_i}$ and $\delta_{\gamma_i}$ are random processes that describe uncertainties in the impedance and propagation constant of the $i$th line standard. Under the assumption that $\delta_{\gamma_i}$ and $\delta_{Z_i}$ are jointly distributed after zero-mean Gaussian distribution, then the joint distribution of $\Gamma_i,\gamma_i$ is also Gaussian, $\Gamma_i,\gamma_i\sim\mathcal{N}\left([\mu_\Gamma,\mu_\gamma],\mat{\Sigma}_{\Gamma_i,\gamma_i}\right)$. Following that $\delta_{Z_i}$ is assumed to be zero-mean Gaussian, then it follows that $\mu_\Gamma=0$. Additionally, for the value of $\mu_\gamma$ we use the estimated value computed by the multiline TRL calibration. The joint covariance $\mat{\Sigma}_{\Gamma_i,\gamma_i}$ is determined through propagating uncertainties in the cross-section of the transmission line (e.g., martial properties and dimensional quantities) by using an appropriate model (e.g., analytical or electromagnetic simulation).

Lastly, to determine $\mat{\Sigma}_\mathrm{I}$, we only need to propagate $\mat{\Sigma}_{\Gamma_i,\gamma_i}$ through the error-box model in (\ref{eq:3.8}), using the estimated values of the calibration coefficients and the propagation constant from the multiline TRL as mean values. Thus, $\mat{\Sigma}_\mathrm{I}$ for the $i$th line standard is given by
\begin{equation}
	\mat{\Sigma}_\mathrm{I}^{(i)} = \mat{J}_{\mat{M}_i^\prime}(\mu_{\Gamma}=0,\mu_{\gamma})\mat{\Sigma}_{\Gamma_i,\gamma_i}\mat{J}^T_{\mat{M}_i^\prime}(\mu_{\Gamma}=0,\mu_{\gamma}).
	\label{eq:3.10}
\end{equation}

\begin{figure}[th!]
	\centering
	\includegraphics[width=1\linewidth]{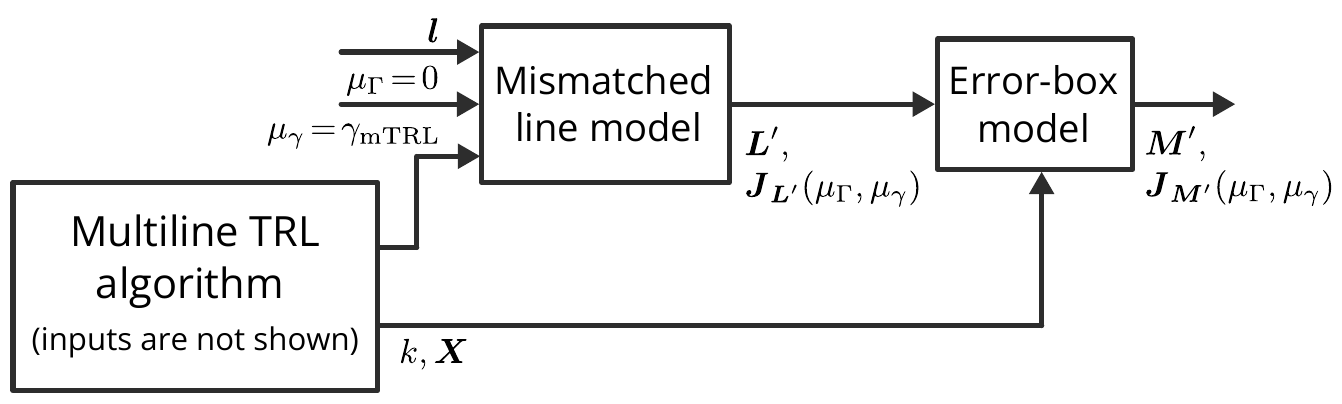}
	\caption{Flow diagram of multiline TRL calibration accounting for line mismatch uncertainty. $\gamma_\mathrm{mTRL}$ and $k\mat{X}$ are the multiline TRL computed propagation constant and calibration coefficients, $\mat{l}$ is a vector containing the length of all line standards, $\mat{L}^\prime$ and $\mat{M}^\prime$ contain the T-parameter of all line standards before and after the application of the error-box model, respectively.}
	\label{fig:3.2}
\end{figure}

With knowledge of $\mat{\Sigma}_\mathrm{I}$, we update the covariance matrix of the measurements and propagate it in the forward uncertainty propagation, as discussed in the previous subsection. The flow diagram in Fig. \ref{fig:3.2} summarizes the steps for inverse uncertainty propagation of the line mismatch.

It is important to note that the definition of uncertainties in the transmission line model can be rephrased in alternative terms. For instance, instead of using the reflection coefficient $\Gamma$, the analysis could be based on the characteristic impedance $Z$, and instead of using the propagation constant $\gamma$, the relative effective permittivity $\epsilon_\mathrm{r,eff}$ could be used. It is also important to remember that the Jacobian matrix needs to be adjusted accordingly.

\section{Experiment}
\label{sec:4}

\begin{figure}[!ht]
	\centering
	\includegraphics[width=0.95\linewidth]{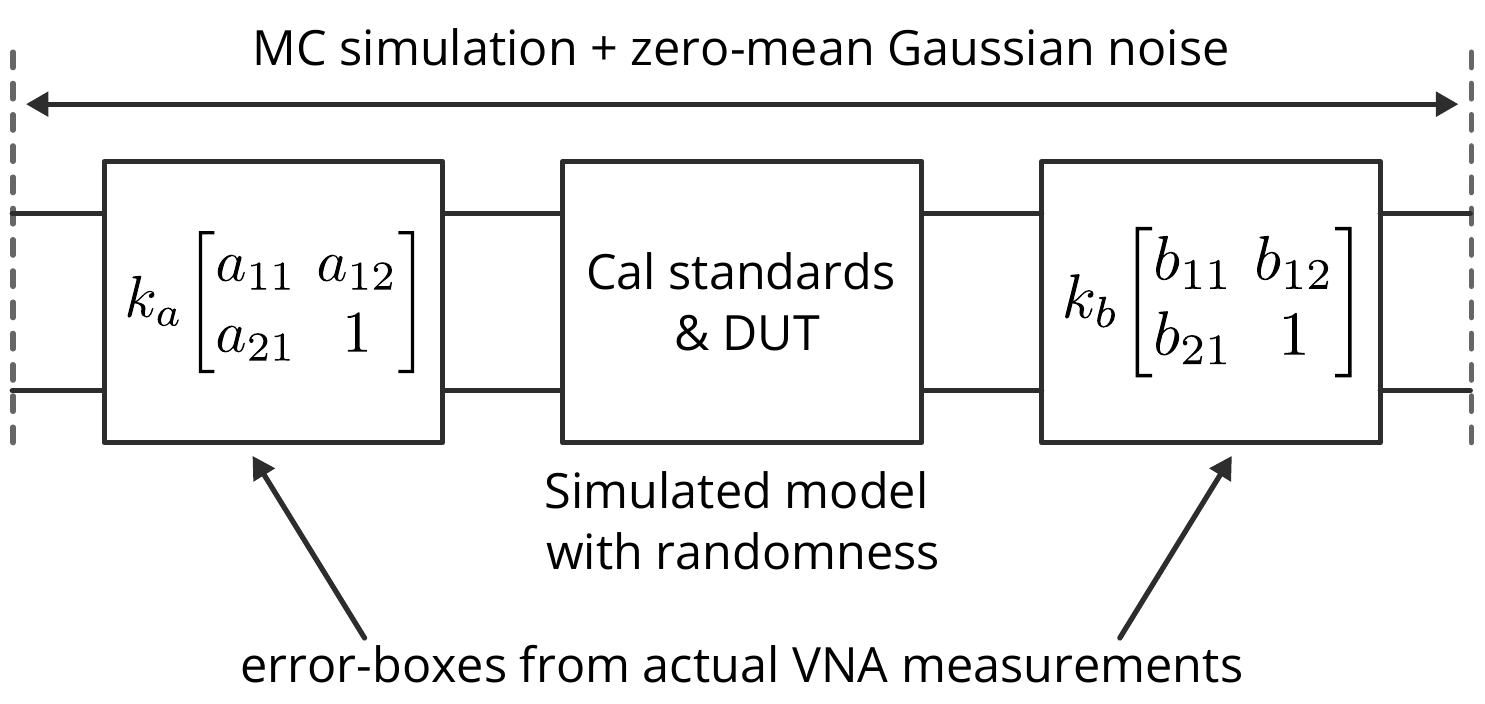}
	\caption{The Experiment setup. The parameters of the calibration standards are perturbed, while Gaussian noise is generated based on the sample covariance estimate obtained from the VNA measurements.}
	\label{fig:4.1}
\end{figure}

To validate our uncertainty propagation technique, we compare it to the MC simulation in which the actual calibration standards are perturbed. For this purpose, we use real measurements to model error-boxes and synthetic data to model coplanar waveguide (CPW) structures as calibration standards. Fig. \ref{fig:4.1} provides an overview of the design of the simulation. In the experiment, the error-boxes remain constant while the parameters of the CPW standards are perturbed by Gaussian noise. Additionally, the entire simulation is perturbed by Gaussian noise based on the covariance matrix estimated from the actual VNA measurements.

\subsection{Measurement}
\begin{figure}[th!]
	\centering
	\includegraphics[width=0.95\linewidth]{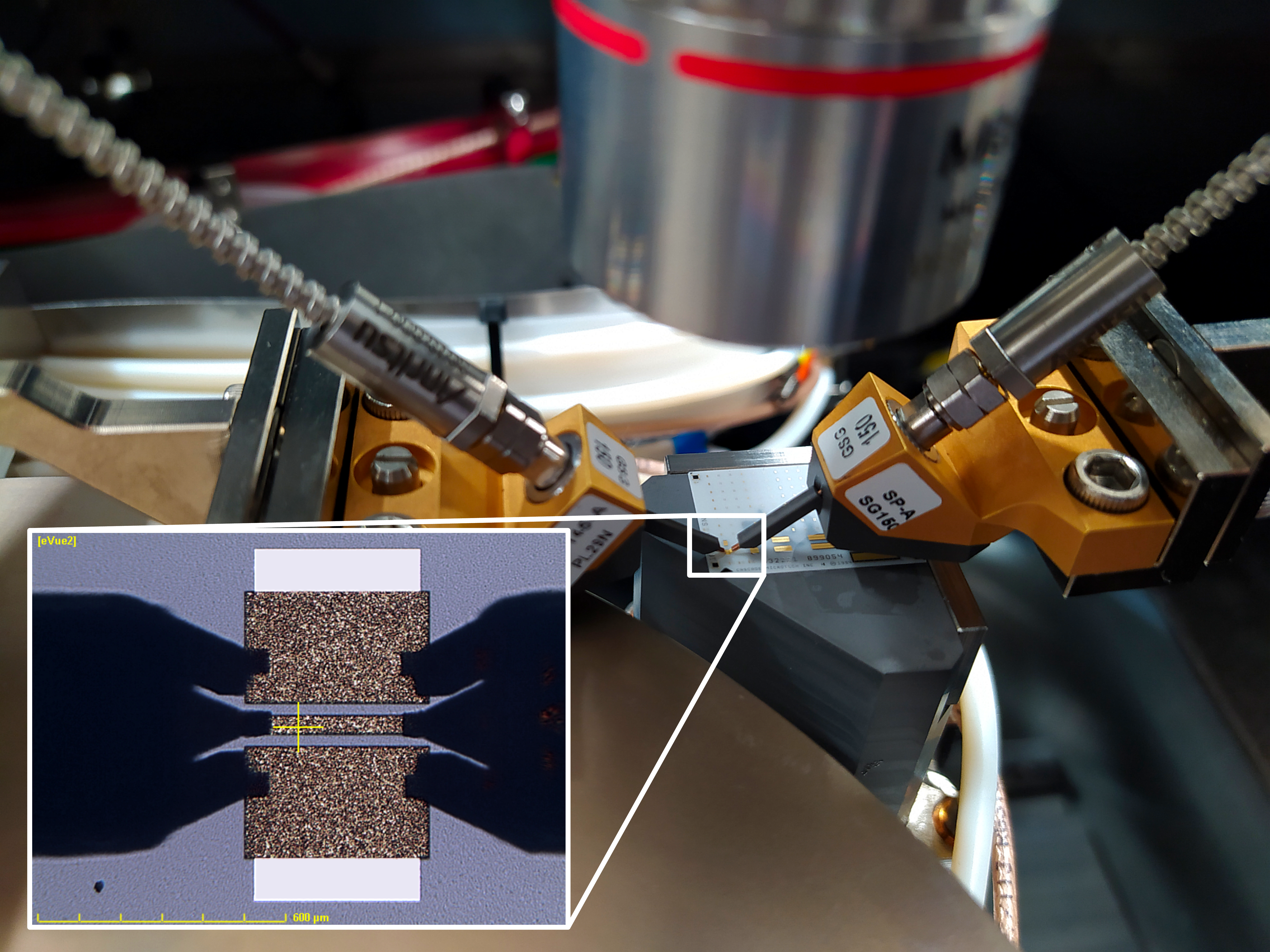}
	\caption{The measurement setup, depicting the ACP probes in contact with the calibration substrate.}
	\label{fig:4.2}
\end{figure}

The measurement setup includes a VectorStar VNA from Anritsu, with mm-wave heads up to 150\,GHz, used together with a SUMMIT200 probe station from FormFactor. The CPW calibration standards were implemented on a commercial impedance standard substrate (ISS) from FormFactor. The measurements were conducted using $150\,\mu\mathrm{m}$-pitch ground-signal-ground (GSG) ACP probes with a 0.8\,mm connector interface from FormFactor. Fig. \ref{fig:4.2} depicts the measurement setup. The calibration standards comprise six CPW lines with edge-to-edge length $\{200,450,900,1800,3500,5250\}\,\mu\mathrm{m}$ and an open standard, which is measured by setting the probes to float.

\begin{figure}[th!]
	\centering
	\subfloat[]{\includegraphics[width=.49\linewidth]{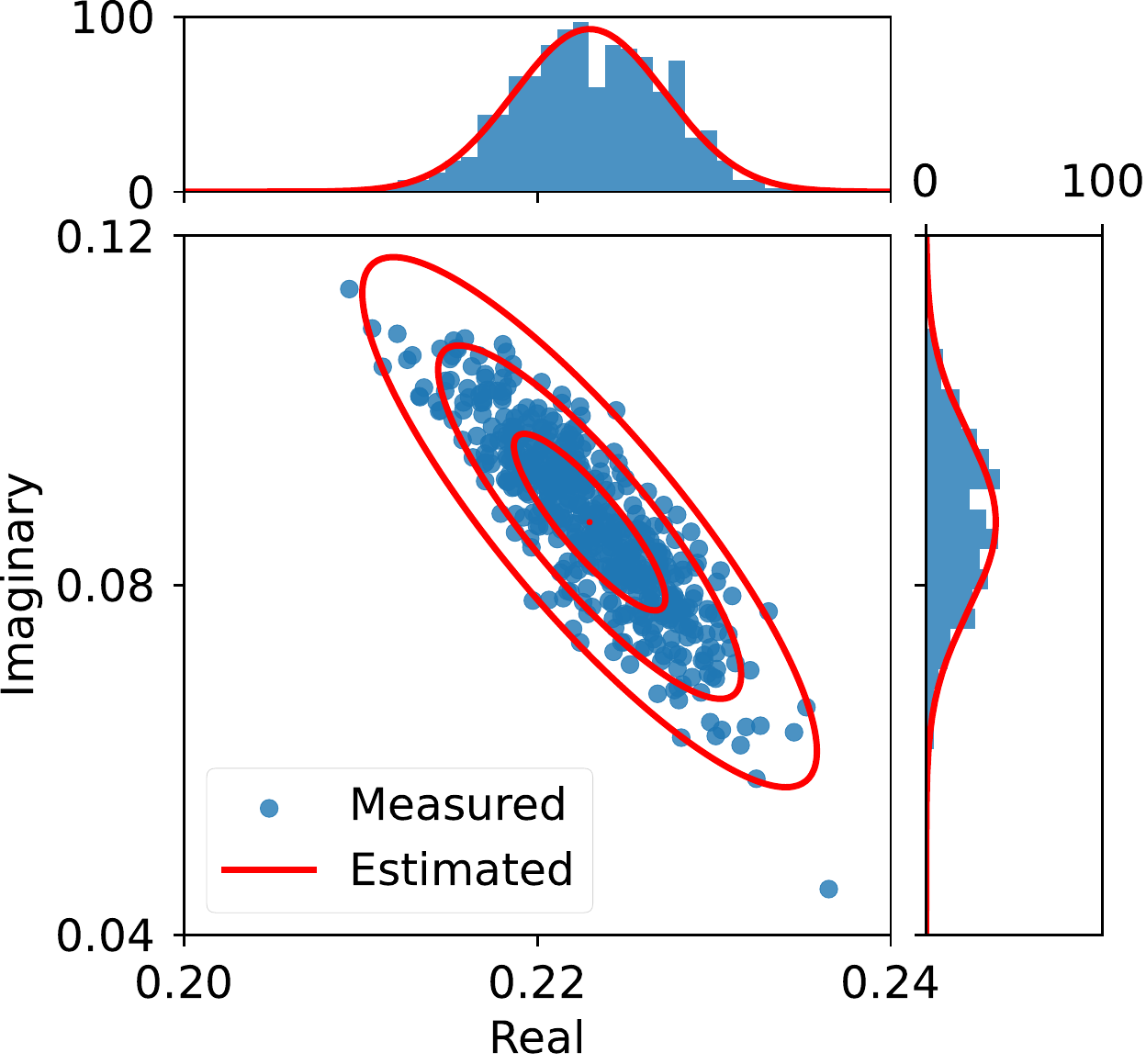}\label{fig:4.2a}}
	\subfloat[]{\includegraphics[width=.49\linewidth]{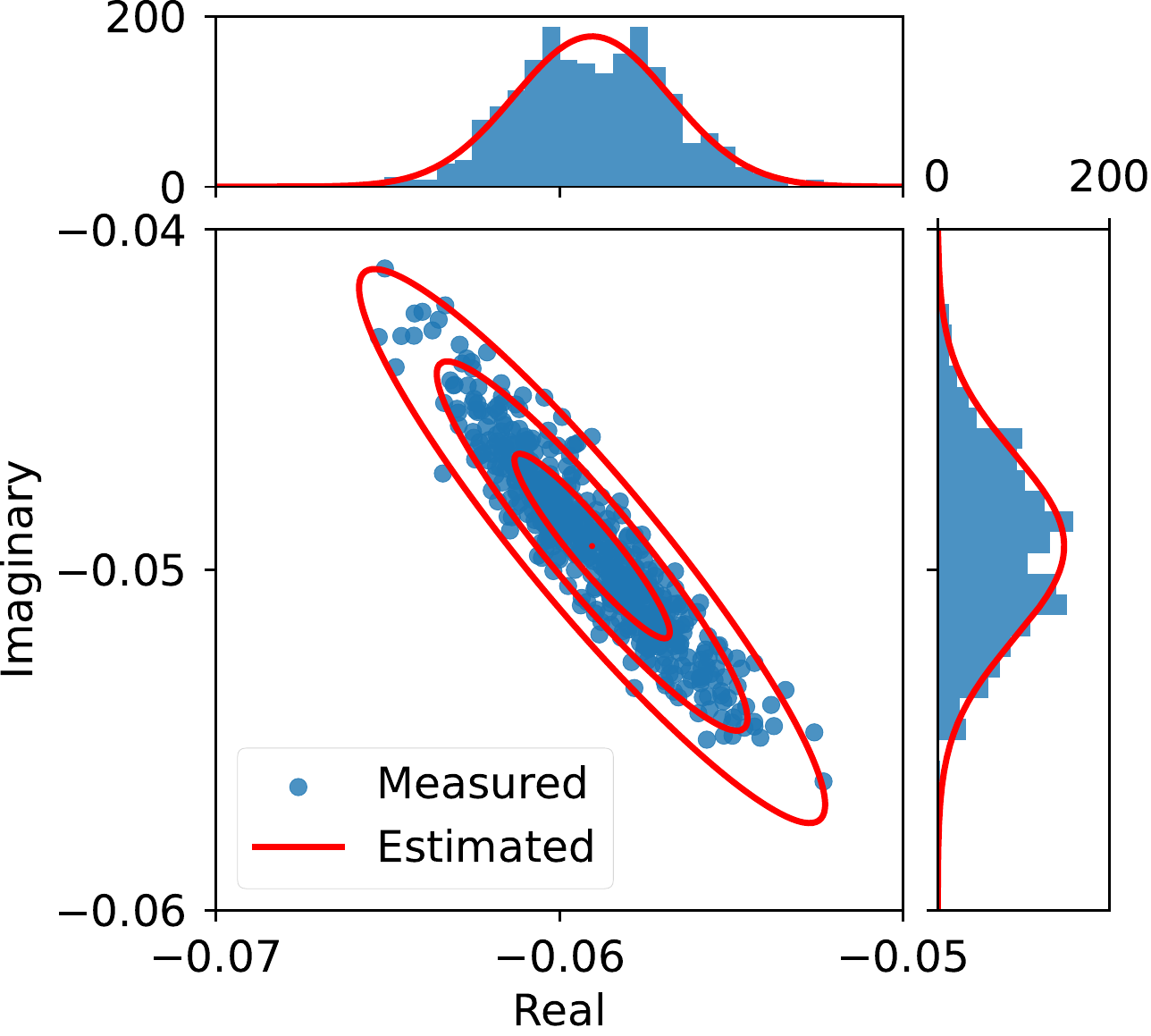}\label{fig:4.2b}}
	\caption{Scatter plots and corresponding marginal histograms from 500 points of uncalibrated measurements of (a) $S_{11}$, (b) $S_{21}$, from the $1800\,\mu\mathrm{m}$ line standard at 120\,GHz. The estimated contour lines of the bivariate Gaussian distribution correspond to a coverage of 68\%, 95\% and 99.7\%, respectively.}
	\label{fig:4.3}
\end{figure}

The measurements were conducted in the frequency range 1 to 150\,GHz with a step of 1\,GHz, resulting in 150 points. A low number of frequency points was necessary to reduce the measurement time, as 500 frequency sweeps of the wave parameters were collected for each standard at an IF-bandwidth of 1\,kHz and power level of -10\,dBm. It took less than 10 minutes to collect all sweeps for each standard, resulting in a measurement time of under 70 minutes for all standards. The S-parameters were calculated and the sample covariance matrix for each standard was estimated. Fig. \ref{fig:4.3} shows scatter plots and the corresponding marginal histograms of the uncalibrated measurements of $S_{11}$ and $S_{21}$ from the $1800\,\mu\mathrm{m}$ line standard at 120\,GHz.


\subsection{CPW cross-section modeling}
To model the CPW structures used in the measurements, we used the dimensional and material parameters along with their uncertainties. Since publicly available data on these parameters are limited, we conducted optical measurements for dimensional parameters, and obtained material properties from publicly available sources. Gold was used for the conductor layer, and lossless Alumina with a relative permittivity of 9.9 was used for the substrate. To demonstrate the propagation of uncertainty in multiline TRL calibration, we assumed $\pm10\%$ standard uncertainty for the smallest planar dimensional feature, $\pm10\%$ for both the thickness and the conductivity of the Gold layer, and $\pm0.2$ for the relative permittivity of the Alumina substrate. These uncertainty assumptions are for demonstration purposes only and are not intended to be considered accurate. Table \ref{tab:4.1} summarizes the parameters and their uncertainties.

\begin{table}[ht!]
	\centering
	\caption{Cross-section dimensions and material properties of the considered CPW structure and their standard uncertainties.}
	\label{tab:4.1}
	\resizebox{\columnwidth}{!}{%
		\begin{tabular}{cccccc}
			\toprule
			\begin{tabular}[c]{@{}c@{}}Signal\\ width\\ ($\mu$m)\end{tabular} &
			\begin{tabular}[c]{@{}c@{}}Ground\\ width\\ ($\mu$m)\end{tabular} &
			\begin{tabular}[c]{@{}c@{}}Conductor\\ spacing\\ ($\mu$m)\end{tabular} &
			\multicolumn{1}{c}{\begin{tabular}[c]{@{}c@{}}Conductor\\ thickness\\ ($\mu$m)\end{tabular}} &
			\begin{tabular}[c]{@{}c@{}}Relative \\ permittivity\\ (1)\end{tabular} &
			\begin{tabular}[c]{@{}c@{}}Conductor\\ conductivity\\ (S/m)\end{tabular} \\ \midrule
			\begin{tabular}[c]{@{}c@{}}$49.1$\\ $\pm2.55$\end{tabular} &
			\begin{tabular}[c]{@{}c@{}}$273.3$\\ $\pm2.55$\end{tabular} &
			\begin{tabular}[c]{@{}c@{}}$25.5$\\ $\pm2.55$\end{tabular} &
			\begin{tabular}[c]{@{}c@{}}$4.9$\\ $\pm0.49$\end{tabular} &
			\begin{tabular}[c]{@{}c@{}}$9.9$\\ $\pm0.2$\end{tabular} &
			\begin{tabular}[c]{@{}c@{}}$4.11\times10^7$\\ $\pm0.41\times10^7$\end{tabular} \\ \bottomrule
		\end{tabular}
	}
\end{table}

\begin{figure}[th!]
	\centering
	\includegraphics[width=0.98\linewidth]{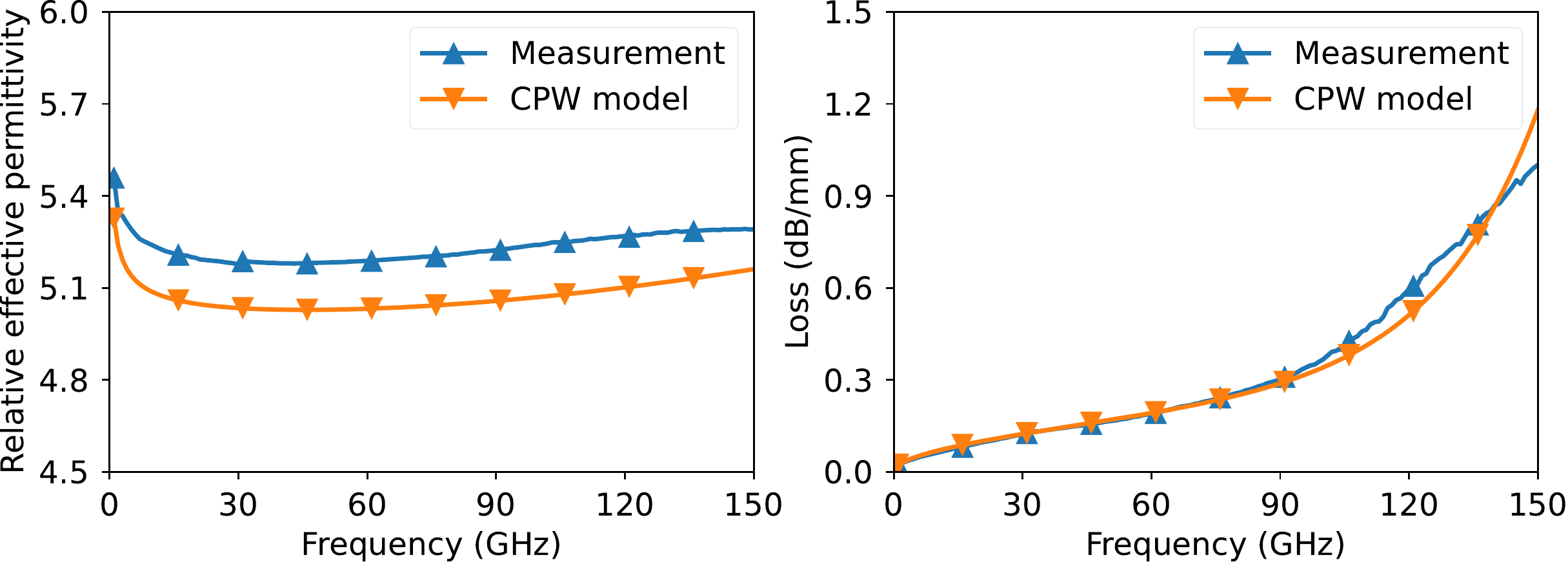}
	\caption{Comparison between measurements and CPW model of the relative effective permittivity and the loss per unit length. The CPW parameters are presented in Table \ref{tab:4.1}.}
	\label{fig:4.4}
\end{figure}

The CPW model we are basing our analysis based on is the analytical model presented in \cite{Phung2021,Schnieder2003,Heinrich1993}. This model includes corrections for radiation loss from references \cite{Phung2021,Schnieder2003}. A comparison between the CPW model based on the data in Table \ref{tab:4.1} and the extracted relative effective permittivity and loss per unit length from the multiline TRL measurements is shown in Fig. \ref{fig:4.4}. The measurements and the model are in reasonable agreement. Therefore, the MC analysis given in the following subsection should represent a realistic scenario of the CPW on-wafer measurements. It is worth noting that the offset seen in the relative effective permittivity can be attributed to the fact that the calibration substrate has a thickness of $254\,\mu\mathrm{m}$ placed on an absorbent chuck, whereas the CPW model assumes an infinite extent of the substrate.

\subsection{MC analysis and linear uncertainty propagation}
The uncertainties considered in our analysis are measurement noise, line mismatch, length offset, and reflect asymmetry. Measurement noise was modeled based on the sample covariance from the measurements. Line mismatch uncertainties were established by perturbing the cross-sectional parameters based on the values in Table \ref{tab:4.1}. The length offset was assumed to have $\pm40\,\mu\mathrm{m}$ standard uncertainty to account for the contact overtravel of the ACP probes. The reflect asymmetry was introduced as an offset of $\pm40\,\mu\mathrm{m}$ standard uncertainty to emulate the effects of changes in the open capacitance. 

For the MC analysis, we generated S-parameter data of the calibration standards using the CPW model and embedded them into the error-boxes obtained from the multiline TRL measurements. We perturbed the relevant parameters in each realization and collected the corresponding output. The uncertainty of the output parameters was estimated as sample covariance. We conducted the MC analysis using 5000 samples to ensure reliable numerical convergence. While it was possible to use more samples, we noted that using as few as 100 samples yielded results comparable to those achieved with 5000 samples. It is worth noting that the MC analysis with 5000 samples took more than an hour and a half to complete.

For linear uncertainty propagation, we estimated the necessary covariance matrices. The covariance for line mismatch was estimated using numerical calculation of the Jacobian matrix of the CPW model, as the analytical model proposed by \cite{Phung2021,Schnieder2003,Heinrich1993} contains the complete elliptic integral of the first, second, and third kind, which hinders the ability to compute analytical derivatives. The covariance of the asymmetry of the open standard was computed analytically by its Jacobian matrix. We implemented the multiline TRL algorithm in Python and used the package \textit{scikit-rf} \cite{Arsenovic2022} for processing S-parameter data and the package \textit{Metas.UncLib} \cite{Zeier_2012} to automatically compute Jacobian matrices and propagate uncertainties linearly.

\begin{figure}[th!]
	\centering
	\includegraphics[width=0.98\linewidth]{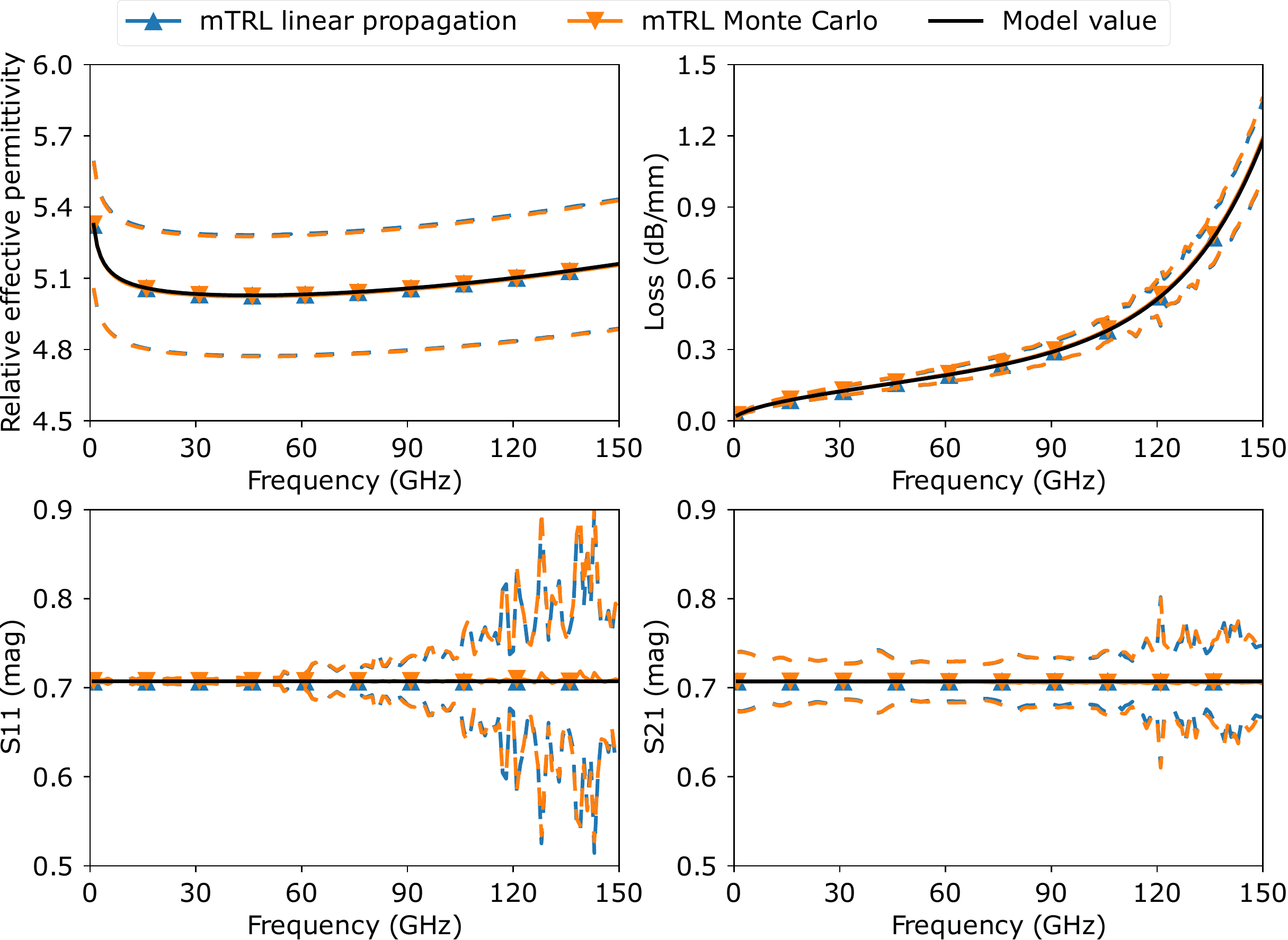}
	\caption{The evaluated relative effective permittivity and loss per unit length from the multiline TRL and the calibrated $S_{11}$ and $S_{21}$ of the DUT. Uncertainty bounds are shown as dashed lines and correspond to the 95\% coverage of the Gaussian distribution. The curve labeled ``Model value'' represents the simulated model before introducing randomness and embedding it with the error boxes.}
	\label{fig:4.5}
\end{figure}

To analyze the impact of uncertainty in calibration, we considered a symmetric, lossless DUT with equal transmission and reflection to be calibrated. In Fig. \ref{fig:4.5}, we compare MC analysis and linear uncertainty propagation for relative effective permittivity and loss per unit length, as well as $S_{11}$ and $S_{21}$ of calibrated DUT. From the plots we can see an excellent agreement between the two methods, validating our approach. Since we are running a finite number of MC trials and the linear propagation remains an approximation, there is an error tolerance between the derived uncertainties from the two methods. On average, over all frequency points, the relative error of the estimated uncertainties from the linear propagation method compared to the MC method is about $0.6\%$ for the relative effective permittivity, $5.33\%$ for the loss per unit length, $4.61\%$ for the magnitude of $S_{11}$, and $4.99\%$ for the magnitude of $S_{21}$. It is worth noting that for the MC method, we perturbed each parameter directly in the simulation, whereas in the linear uncertainty propagation, we only provided the covariance matrices. This means that if the covariance matrices are available (even if it is a rough estimate), the uncertainty evaluation can be performed quickly, simultaneously with the calibration.

\begin{figure}[th!]
	\centering
	\includegraphics[width=0.98\linewidth]{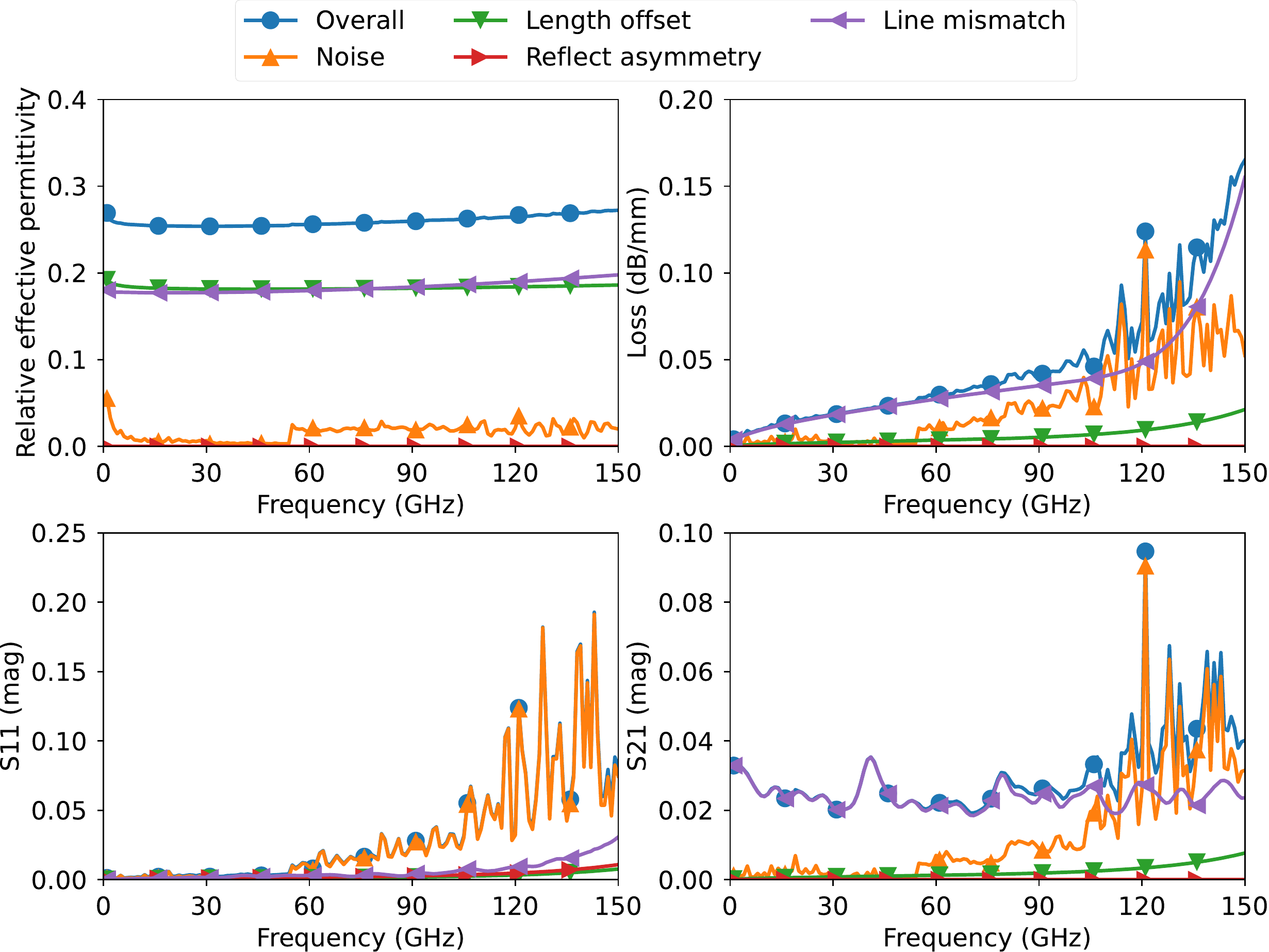}
	\caption{The uncertainty budget due to the individual uncertainty types. The uncertainties are given as 95\% coverage of the Gaussian distribution.}
	\label{fig:4.6}
\end{figure}

Linear uncertainty propagation allows for a quick evaluation of the uncertainty budget caused by different sources of uncertainty. As shown in Fig. \ref{fig:4.6}, the impact of each type of uncertainty can be analyzed. For example, length uncertainty and line mismatch have significant uncertainties in the extracted relative effective permittivity. In addition, noise in measurements affects all quantities significantly, except for the relative effective permittivity. As observed in the uncertainty budget, the noise contribution from the VNA is higher than expected at higher frequencies, most significantly after 100 GHz. After contacting the VNA manufacturer, we discovered that the issue originated from additional statistical error introduced by the VNA firmware. The manufacturer was able to fix this issue with an updated firmware version. This experience emphasizes the usefulness of analyzing the uncertainty budget, as it allows for proper debugging.

\begin{figure}[th!]
	\centering
	\includegraphics[width=0.98\linewidth]{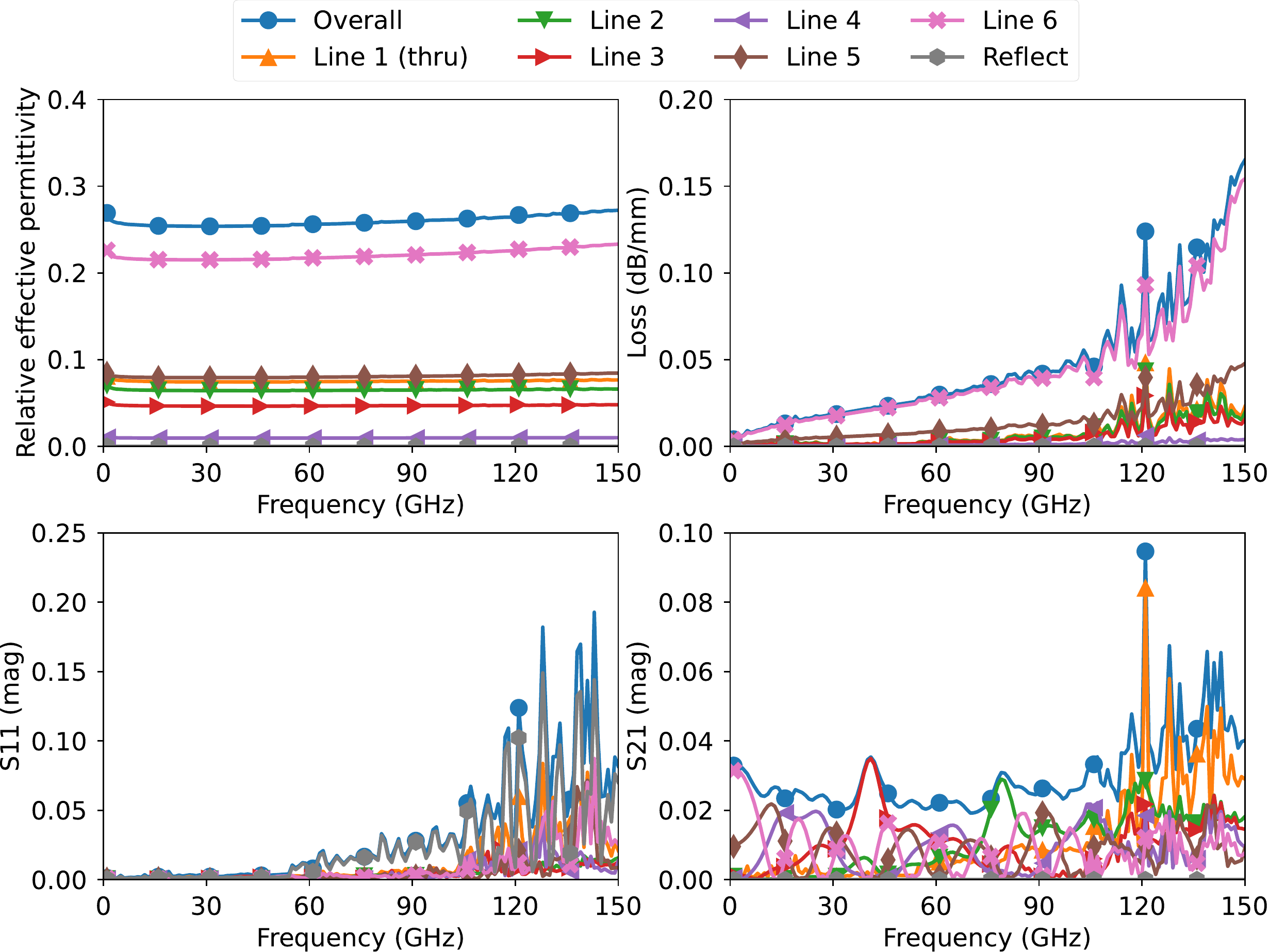}
	\caption{The uncertainty budget due to the individual calibration standards. The uncertainties are given as 95\% coverage of the Gaussian distribution.}
	\label{fig:4.7}
\end{figure}

Furthermore, by breaking down the uncertainty contribution, we can see how each standard contributes to the overall uncertainty, as shown in Fig. \ref{fig:4.7}, which can be used to optimize calibration standards and identify measurement bottlenecks.

\section{Conclusion}
\label{sec:5}

In summary, we presented an efficient method for linear uncertainty propagation in multiline TRL calibration. This approach allows for a quick and easy evaluation of the uncertainty budget caused by different sources of uncertainty, such as measurement noise, line mismatch, length uncertainty, and reflect asymmetry. The results were validated by comparing them to those obtained from an MC analysis. It is important to note that the proposed method is based on the principles of the ISO GUM, which includes assumptions such as linearization and a Gaussian distribution of the uncertainties. These assumptions should be considered when interpreting the results, as they may limit the scope of the method’s applicability. Despite these limitations, the presented approach still provides valuable insights into the uncertainty contributions within the calibration. It can give a quick overview of the primary sources of uncertainty in the measurements.


\section*{Acknowledgment}
The financial support by the Austrian Federal Ministry for Digital and Economic Affairs and the National Foundation for Research, Technology, and Development is gratefully acknowledged. The authors also thank ebsCENTER for providing access to their measurement equipment, Martin Medebach for helping with programming the VNA, and Sitaram Stepponat for helping with measuring the dimensions of the calibration substrate.

\bibliographystyle{IEEEtran}
\bibliography{References/references.bib}

\end{document}